\documentclass[12pt,floatfix,showpacs,preprintnumbers,amssymb]{iopart}

\usepackage{graphicx}
\usepackage{pifont}

\graphicspath{{fig/}}
\usepackage{dcolumn}
\usepackage{bm}
\usepackage{float}
\usepackage{amssymb}
\usepackage{color}

\usepackage{textcomp}

\begin{document}

\title[Temperature dependence of  quantum spin pumping  ]
{The temperature dependence of quantum
spin pumping generated using electron
spin resonance with three-magnon
splittings
}

\author{Kouki Nakata}

\address{Yukawa Institute for Theoretical Physics, 
Kyoto University,  \\
Kitashirakawa Oiwake-Cho, Kyoto 606-8502, Japan}
\ead{nakata@yukawa.kyoto-u.ac.jp}

\begin{abstract}
On the basis of the Schwinger-Keldysh formalism,
we have closely investigated the temperature dependence of quantum spin pumping  by electron spin resonance. 
We have clarified that
three-magnon splittings excite  non-zero modes of magnons and 
characterize the temperature dependence of quantum spin pumping.
Our theoretical result qualitatively agrees with the experiment by Czeschka et al. 
that
the mixing conductance is little influenced by temperature
[F. D. Czeschka et al., Phys. Rev. Lett., $\mathbf{107}$, 046601 (2011)].
\end{abstract}

\noindent{\it magnon, quantum spin pumping, electron spin resonance, three-magnon splittings, spintronics   \/}:
\submitto{J. Phys.: Condens. Matter}
\noindent{YITP-13-6}
\maketitle

\section{Introduction}
\label{sec:intro}

The standard way to generate spin currents is spin pumping,\cite{uchidaTSP,sandweg,mod2,bauerreview}
which has already been established experimental technique at finite temperature;\cite{saitohprivate,ISHE1}
Czeschka et al.\cite{czeschka} have experimentally showed that
the mixing conductance is little influenced by temperature.
On top of this,
some of technologies of spintronics are in fact going to be put into practical use;\cite{saitohprivate}
they are applied to green information and communication technologies.
In contrast to such experimental development,  to the best of our knowledge,
theoretical studies so far of quantum spin pumping at finite temperature beyond phenomenologies, unfortunately,  
will not be enough to explain the experimental result,
in particular the above one by Czeschka et al.\cite{czeschka} 
Hence for the further development of spintronics and the application,
the theoretically rigorous description of spin pumping at finite temperature
is an urgent and important subject.

In order to overcome such a theoretical situation,
we employ the Schwinger-Keldysh formalism\cite{rammer,kita,tatara,haug,kamenev,new}
and clarify the features of quantum spin pumping at finite temperature.
This is the main aim of this paper.
Let us remark that
for the experimental realization of spin pumping,
the time-dependent transverse magnetic field, which acts as  `quantum fluctuations',\cite{oshikawaprivate,altland}
is applied and it drives the system into a non-equilibrium steady state.
For the theoretical description of such systems beyond phenomenologies,
one of the most suitable theoretical tools will be the Schwinger-Keldysh formalism;\cite{rammer,kita,tatara,haug,kamenev,new}
owing to the Schwinger-Keldysh closed time path,\cite{QSP,nakatatatara,tataraprivate,rammer}
this  formalism is not based on the assumption called the (well-known) 
Gell-Mann and Low theorem.\cite{kamenev,fetter,peskin,Gell-mann,kita}
Therefore within the perturbative theory, 
the formalism can deal with an arbitrary time-dependent Hamiltonian\cite{tataraprivate}
and can treat the system out of the equilibrium.
On top of this,
the Schwinger-Keldysh formalism is applicable to systems at finite temperature;
the well-known Matsubara formalism\cite{matsubara} (i.e. the imaginary-time formalism),\cite{fetter}
which also can deal with thermodynamic average value,
can be regarded as a simple corollary of the Schwinger-Keldysh formalism 
(i.e. closed time path formalism or the real-time formalism).\cite{rammer}
That is, the Schwinger-Keldysh formalism includes the  Matsubara formalism
and  information about finite temperature is contained in the greater and lesser Green's functions.\cite{tataraprivate} 
Consequently we can treat non-equilibrium phenomena at finite temperature
owing to the Schwinger-Keldysh formalism.

\begin{figure}[b]
\begin{center}
\includegraphics[width=7cm,clip]{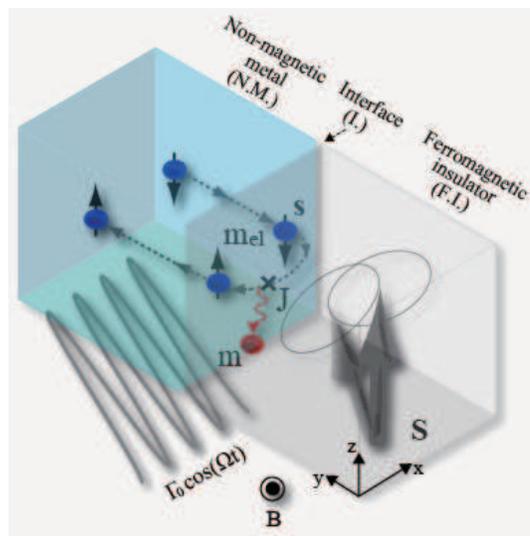}
\caption{
The schematic picture of the quantum spin pumping system.
 Spheres  represent magnons and those with arrows are conduction electrons.
 The wavy line denotes the time-dependent transverse magnetic field $\Gamma (t)$ (i.e. the external pumping field).
The interface is defined as an effective area where the Fermi gas (conduction electrons)  and the Bose gas (magnons) coexist to interact; $J\not=0$.
The width of the interface might be supposed to be of the order of the lattice constant.\cite{interface}
The interface can be regarded also as a ferromagnetic metal.\cite{Bauer}
Conduction electrons  cannot  enter the ferromagnet, which is an insulator. 
{\textit{Note; clear pictures are available at the following URL;}} 
https://dl.dropbox.com/u/5407955/JPhysCondMatQSPtemp.pdf
 \label{fig:pumping} }
\end{center}
\end{figure}

Actually in our previous work,\cite{QSP}
we have already reformulated the quantum spin pumping theory
from the viewpoint of the Schwinger-Keldysh formalism
and have shown that spin pumping can be generated also by electron spin resonance (ESR)\cite{QSP} 
as well as ferromagnetic resonance (FMR);\cite{mod2,sandweg,ISHE1,battery,nakatatatara}
this is the natural result from the fact that
the applied time-dependent transverse magnetic field (i.e. quantum fluctuations)
affects conduction electrons as well as localized spins (i.e. magnons) at the interface (Fig. \ref{fig:pumping}).
To clarify the temperature dependence of quantum spin pumping  by ESR\cite{QSP}
and find the microscopic origin is the final goal of this paper.

The quantum spin pumping system we had treated reads as follows;\cite{QSP}  
we consider a ferromagnetic insulator and  non-magnetic   metal   junction shown in Fig. \ref{fig:pumping}.
At the interface,
conduction electrons couple with localized spins  $  {\mathbf{S}}({\mathbf{x}},t) $, ${\mathbf{x}}=(x, y, z) \in {\mathbb{R}}^3$;
\begin{equation}
 {\cal{H}}_{\rm{ex}}=  - 2Ja_0^3  {\int_{{\mathbf{x}}\in \rm{(interface)}}} d {\mathbf{x}}   {\ }  {\mathbf {S}}({\mathbf{x}},t) \cdot   {\mathbf {s}}({\mathbf{x}},t),
 \label{eqn:qsp1}
\end{equation}
where the lattice constant of the ferromagnet is denoted as $a_0$.
The magnitude of the interaction is supposed to be constant
and it is expressed  as  $ 2J $. 
Note that
owing to this exchange interaction $  {\cal{H}}_{\rm{ex}}$ at the interface,
the spin angular momentum can be interchanged between conduction electrons and the ferromagnet.
That is, this exchange interaction $  {\cal{H}}_{\rm{ex}}$ at the interface is the key to spin pumping.\cite{saitohprivate}
Therefore we identify the system characterized by the exchange interaction between conduction electrons and the ferromagnet $  {\cal{H}}_{\rm{ex}}$ 
(Hamiltonian (\ref{eqn:qsp1})) with the spin pumping system.
From now on,
we exclusively focus on the dynamics at the interface.

Conduction electron spin variables are represented as
$ {{s}^{j}}  = \sum_{\eta ,\zeta  = \uparrow , \downarrow} c^{\dagger }_{\eta} (\sigma  ^j)_{\eta \zeta} c_{\zeta}/2    
               =:  c^{\dagger }  \sigma ^j c/2$
with the $ 2\times 2 $  Pauli matrices;
$ [ \sigma  ^j , \sigma  ^k  ] = 2i\epsilon _{jkl} \sigma  ^l   $, ($ j, k, l = x,y,z$).
Operators $c^{\dagger }/c $ are   creation/annihilation operators for conduction electrons
satisfying the (fermionic) anticommutation relation;
$ \{c_{\eta  }({\mathbf{x}}, t), c_{\zeta }^{\dagger }({\mathbf{x}}', t) \}= \delta _{\eta , \zeta } \delta ({\mathbf{x}}-{\mathbf{x'}})  $.
We suppose the uniform magnetization and
thus,  localized spin degrees of freedom   can be mapped into magnon\cite{sandweg,spinwave,kurebayashi} ones 
via the Holstein-Primakoff transformation;
\begin{eqnarray}
S^+({\mathbf{x}},t)
&=& \sqrt{2\tilde S}  \sqrt{1-\frac{a^\dagger({\mathbf{x}},t) a({\mathbf{x}},t)}{2\tilde S}} a({\mathbf{x}},t)   \label{eqn:qsp3-1}  \\
&=& \sqrt{2\tilde S} \Big[1- \frac{a^\dagger({\mathbf{x}},t) a({\mathbf{x}},t)}{4\tilde S} \Big]a({\mathbf{x}},t)    
+{\cal{O}}({\tilde S}^{-3/2}) \label{eqn:qsp3-2}  \\
&=&  (S^{-})^{\dagger }  \label{eqn:qsp3-2-2}   \\
S^z({\mathbf{x}},t) 
&=& \tilde S-a^\dagger ({\mathbf{x}},t)  a   ({\mathbf{x}},t),
 \label{eqn:qsp3}
\end{eqnarray}        
with $\tilde S  :=  S/{a_0^3}$.
Operators $a^{\dagger }/a $ are magnon creation/annihilation operators
satisfying the (bosonic) commutation relation; 
$ [a(\mathbf{x}, t), a^{\dagger }(\mathbf{x'}, t) ]= \delta (\mathbf{x}-\mathbf{x'}) $.
Up to the $ {\cal{O}}(\tilde S)$ terms,
localized spins are reduced to  free boson degrees of freedoms.
Consequently, in the quadratic dispersion (i.e. long wavelength) approximation, 
the dynamics of localized spins  with the applied magnetic field along the quantization axis (i.e. z-axis) $B$  is described
by the Hamiltonian $  {\cal{H}}_{\rm{mag}} $;
\begin{eqnarray}
  {\cal{H}}_{\rm{mag}}  =  \int_{{\mathbf{x}}\in \rm{(interface)}} d   {\mathbf{x}}  {\ }  {a^{\dagger }(\mathbf{x},t)} 
                                             \Big(-\frac{ { \mathbf{\nabla }}^2 }{2m} + B \Big)  
                                             {a(\mathbf{x},t)},
\label{eqn:qsp01}
\end{eqnarray}
where the effective mass of magnons is denoted by $m$.
In addition, Hamiltonian  $ {\cal{H}}_{\rm{ex}}$ can be rewritten as 
${\cal{H}}_{\rm{ex}} =     {\cal{H}}_{\rm{ex}} ^{S}+{\cal{H}}_{\rm{ex}} ^{\prime}$ with
\begin{eqnarray}
 {\cal{H}}_{\rm{ex}} ^{S}   &=&   
-JS  {\int_{{\mathbf{x}}\in \rm{(interface)}}} d    {\mathbf{x}}  
 {\ }   c^{\dagger } ({\mathbf{x}},t)  \sigma^z  c  ({\mathbf{x}},t),        \label{eqn:qsp4-2}   \\                                                                                            
{\cal{H}}_{\rm{ex}} ^{\prime} &=&   -Ja_0^3  \sqrt{\frac{\tilde S}{2}} {\int_{{\mathbf{x}}\in \rm{(interface)}}} d   {\mathbf{x}}  
                                                       \Bigg\{ a^{\dagger }({\mathbf{x}},t) \Big[1-\frac{a^{\dagger }({\mathbf{x}},t) a({\mathbf{x}},t) }{4\tilde S} \Big] 
                                                                                         c^{\dagger }  ({\mathbf{x}},t) \sigma ^{+} c({\mathbf{x}},t )   \nonumber  \\
                                                     &+ &            \Big[1-\frac{a^{\dagger }({\mathbf{x}},t) a({\mathbf{x}},t) }{4\tilde S} \Big] a({\mathbf{x}},t)  
                                                                                                 c^{\dagger } ({\mathbf{x}},t)  \sigma ^{-} c({\mathbf{x}},t ) \Bigg\}    
                                                     + {\cal{O}}({\tilde S}^{-3/2}).
 \label{eqn:qsp4}
\end{eqnarray}
Note that we have adsorbed the Bohr magneton and the $g$-factors into the definition of the magnetic field $B$
and have taken $\hbar =1$ for convenience. 

Here it should be stressed that according to Hamiltonian (\ref{eqn:qsp4-2}),
 the localized spin $S$ acts as an effective magnetic field along the quantization axis for conduction electrons;
\begin{eqnarray}
 {\cal{H}}_{\rm{ex}} ^{S}   &=&   
-JS  {\int_{{\mathbf{x}}\in \rm{(interface)}}} d    {\mathbf{x}}  
 {\ }   c^{\dagger } ({\mathbf{x}},t)  \sigma^z  c  ({\mathbf{x}},t)      \\
 &=& -2JS  {\int_{{\mathbf{x}}\in \rm{(interface)}}} d    {\mathbf{x}}   \          s^z ({\mathbf{x}},t).
 \label{eqn:qsp4-3}
\end{eqnarray}
Therefore,
the diagonal part of the conduction electrons is written as
\begin{eqnarray}
{\cal{H}}_{\rm{el}} &= &\int_{{\mathbf{x}}\in \rm{(interface)}} d   {\mathbf{x}}   {\ }  c^{\dagger } ({\mathbf{x}},t) 
                                       \Big[  -\frac{\nabla ^2}{2m_{\rm{el}}}  -(JS + \frac{B}{2} )\sigma^z   \Big] c({\mathbf{x}},t),
 \label{eqn:qsp5}
\end{eqnarray}
with the effective mass of the conduction electron  $  m_{\rm{el}}$.

For the experimental realization of spin pumping,\cite{sandweg,AndoPumping,ISHE1}
a time-dependent transverse magnetic field $ \Gamma (t) $ with an angular frequency $\Omega $ is applied into the system
as a driving energy; $  \Gamma (t) :=  \Gamma _0 {\rm{cos}}(\Omega t)   $.
This applied periodic transverse magnetic field acts as  `quantum fluctuations'\cite{oshikawaprivate,altland}
and drives the system into a non-equilibrium steady state.\cite{bunkov}
Thus we identify the system described by the exchange interaction $  {\cal{H}}_{\rm{ex}}$  (Hamiltonian (\ref{eqn:qsp1})) 
under the applied time-dependent transverse magnetic field $ \Gamma (t)$
with the `quantum spin pumping system'.
 Note that 
the applied time-dependent transverse magnetic field couples with conduction electrons as well as localized spins;
\begin{eqnarray}
 V_{\rm{el}}^{\Gamma }&=& \frac{\Gamma (t)}{4} \int_{{\mathbf{x}}\in \rm{(interface)}} d   {\mathbf{x}}   {\ } 
                                                 c^{\dagger } ({\mathbf{x}},t) ( \sigma^{+}  +  \sigma^{-}  ) c({\mathbf{x}},t)   \label{eqn:qsp7-2}   \\
 V_{\rm{mag}}^{\Gamma }&= &\Gamma (t)\sqrt{\frac{\tilde S}{2}} \int_{{\mathbf{x}}\in \rm{(interface)}}  d   {\mathbf{x}}    {\ } \Bigg\{ 
                                                        \Big[1-\frac{a^{\dagger }({\mathbf{x}},t) a({\mathbf{x}},t) }{4\tilde S} \Big] a({\mathbf{x}},t)    \nonumber      \\
                                                &+&a^{\dagger }({\mathbf{x}},t) \Big[1-\frac{a^{\dagger }({\mathbf{x}},t) a({\mathbf{x}},t) }{4\tilde S} \Big]    \Bigg\}.
 \label{eqn:qsp7}
\end{eqnarray}
Therefore spin pumping can be generated also by ESR ($\Omega = 2JS +B $)\cite{QSP} as well as FMR ($ \Omega =B$).

Finally,
 the total Hamiltonian of the quantum spin pumping system ${\cal{H}}$
(i.e. the spin pumping system with $\Gamma (t)$) is arranged as
\begin{eqnarray}
{\cal{H}}  :=   { \cal{H}}_{\rm{mag}} +  { \cal{H}}_{\rm{ex}}^{\prime} +  { \cal{H}}_{\rm{el}} + V_{\rm{el}}^{\Gamma }+  V_{\rm{mag}}^{\Gamma }.
 \label{eqn:qsp8}
\end{eqnarray}
We investigate the features of quantum spin pumping described by this Hamiltonian  (Hamiltonian (\ref{eqn:qsp8})); 
we clarify the behavior of quantum spin pumping generated by ESR at finite temperature  and go after the microscopic origin.
This is the main aim of this paper.

The remain of the paper is organized as follows;
we quickly review our quantum spin pumping theory\cite{QSP} and stress the point in sec. \ref{sec:theory}.
The readers who are familiar with our formalism can skip sec. \ref{sec:theory}.
The temperature dependence of quantum spin pumping by ESR is revealed in sec. \ref{sec:temp}.
We go after the microscopic origin and qualitatively understand the behavior from the viewpoint of three-magnon splittings. 

\section{Quantum spin pumping theory based on  
                        Schwinger-Keldysh formalism}
\label{sec:theory}

Before getting straight to the explanation of our quantum spin pumping theory,
let us remark  a point.
In the last section,
localized spin degrees of freedom have been mapped into magnon ones via the Holstein-Primakoff transformation;
we have expanded it up to the ${\cal{O}}({\tilde S}^{-1/2})$ terms (see eqs. (\ref{eqn:qsp3-2}) and  (\ref{eqn:qsp3-2-2})).
Therefore the magnon-magnon interaction, $a^{\dagger }a^{\dagger }aa = {\cal{O}}({\tilde S}^0)$, 
and the magnon-electron interaction, $  a^{\dagger }a c^{\dagger } \sigma ^z c = {\cal{O}}({\tilde S}^0) $,
may emerge as well as Hamiltonian  (\ref{eqn:qsp01}), (\ref{eqn:qsp4-2}) and (\ref{eqn:qsp8}).
Nevertheless, we have omitted the terms.
The reason reads as follows;
in sharp contrast to  $ {\cal{H}}_{\rm{ex}} ^{\prime} $, $ V_{\rm{mag}}^{\Gamma } $ and $ V_{\rm{el}}^{\Gamma }$, 
 (see Hamiltonian (\ref{eqn:qsp4}),  (\ref{eqn:qsp7}) and (\ref{eqn:qsp7-2})),
the magnon-magnon interaction and the magnon-electron interaction do not include the spin-flip operators ($\sigma ^{\pm } $)\footnote{
As the result, they commute with the z-component of the spin density $ \rho_{\rm{s}}^z := {c^{\dagger } \sigma ^z c}/{2}$ and hence, 
they do not directly contribute to the SRT defined in eq. (\ref{eqn:qsp11}).
}
and on top of this, they  consist of the even number in respect to magnon creation/annihilation operators.\footnote{
Now, we have focused on the SRT accompanied by the exchange interaction $J$ between conduction electrons and magnons 
(eqs. (\ref{eqn:qsp10}) and  (\ref{eqn:qsp18})). 
Although $V_{\rm{mag}}^{\Gamma} $ does not  include spin-flip operators ($\sigma^{\pm }$), 
it consists of the odd number in respect to magnon creation/annihilation operators and hence, 
the term $V_{\rm{mag}}^{\Gamma} $ is essential to spin pumping; 
from the viewpoint of the calculation based on the perturbation theory (i.e. Wick's theorem), 
one can easily see that it directly contributes to the SRT represented by eq. (\ref{eqn:qsp10})  (i.e. spin pumping). 
In other words, it is clear, from the viewpoint of Wick's theorem,
that the SRT becomes zero without $V_{\rm{mag}}^{\Gamma} $.
}
Consequently the terms can not directly contribute to spin pumping;\cite{leggett}
as shown in Fig. \ref{fig:pumping}, spin-flip processes described by   spin-flip operators $\sigma ^{\pm } $ are essential to spin pumping.
Therefore, within the approximation of eq. (\ref{eqn:qsp18}), 
we are allowed to omit the effects of the magnon-magnon interaction and the magnon-electron interaction.\footnote{
The magnon-magnon interaction and the magnon-electron interaction indirectly lead to higher-terms than eq. (\ref{eqn:qsp18}),
which are out of the purpose of the present work;
they give  $ {\cal{O}  }(J^2)$-terms for instance.
}

\subsection{Breaking of  spin conservation law }
\label{subsec:break}

We have formulated the spin pumping theory on the basis of the spin continuity equation for conduction electrons;\cite{QSP,tataraprivate,TSP}
 \begin{eqnarray}
 \dot \rho_{\rm{s}}^z  +   \nabla \cdot {\mathbf{j}}_{\rm{s}}^{z} =  {\mathcal{T}}_{\rm{s}}^z,
 \label{eqn:qsp11}
 \end{eqnarray}
where the dot denotes the time derivative of the z-component for the spin density defined as
$  \rho_{\rm{s}}^z :=   {c^{\dagger }  \sigma ^z c}/{2}$, 
 and  ${\mathbf{j}}_{\rm{s}}$  is the spin current density.
 Let us emphasize that in sharp contrast to the case of charges,
the spin conservation law is broken and it is represented by the spin relaxation torque (SRT)\cite{QSP,TSP} ${\mathcal{T}}_{\rm{s}} $,
which appears in the spin continuity equation.
Through the Heisenberg equation of motion,
the  z-component of the SRT can be explicitly written down as 
\begin{eqnarray}
{\mathcal{T}}_{\rm{s}}^z  &=&     iJa_0^3  \sqrt{\frac{\tilde S}{2}}  \Big{\{} a^{\dagger }({\mathbf{x}},t) 
                                                         \Big[1-\frac{ a^{\dagger }({\mathbf{x}},t) a({\mathbf{x}},t)}{4 \tilde S }\Big]      c^{\dagger }  ({\mathbf{x}},t) \sigma^{+}   c({\mathbf{x}},t ) \nonumber \\
                                             & - &  \Big[1-\frac{ a^{\dagger }({\mathbf{x}},t) a({\mathbf{x}},t)}{4 \tilde S }\Big]a({\mathbf{x}},t)  
                                                          c^{\dagger } ({\mathbf{x}},t)  \sigma^{-}   c ({\mathbf{x}},t)  \Big{\}}    \nonumber  \\
                                             &+ & \frac{\Gamma (t)}{4i}  
                                                       \Big[ c^{\dagger }   ({\mathbf{x}},t) \sigma^{+}   c({\mathbf{x}},t )    -    c^{\dagger } ({\mathbf{x}},t)  \sigma^{-}   c ({\mathbf{x}},t) \Big].  
 \label{eqn:qsp10}
\end{eqnarray}
Note that the SRT has arisen from $ { \cal{H}}_{\rm{ex}}^{\prime}$  and  $  V_{\rm{el}}^{\Gamma } $,
which consist of spin-flip operators ($ \sigma^{\pm }$) and quantum fluctuations ($ \Gamma (t)$);
 ${\mathcal{T}}_{\rm{s}}^z  ={[\rho_{\rm{s}}^z,  { \cal{H}}_{\rm{ex}}^{\prime}    +  V_{\rm{el}}^{\Gamma }  ]}/{i}$.
Let me emphasize that 
though each spin conservation law for conduction electrons and magnons is broken,
the total spin angular momentum  is  conserved
in the spin pumping systems.\cite{QSP,TSP}

\subsection{Pumped net spin current}
\label{subsec:net}

One can easily see that
the expectation value of the spin density for conduction electrons reads
$  \langle  {{  \rho_{\rm{s}}^z}  } \rangle  = \sum_{n=0, \pm 1}  \langle {{  \rho_{\rm{s}}^z}  }(n) \rangle  {\rm{e}}^{2in\Omega t} + {\cal{O}}(\Gamma ^4) $,
where $  {{  \rho_{\rm{s}}^z}  }(n) $ represents the (time-independent) expansion coefficient of each angular frequency mode.
Thus
the time-average of the time-derivative becomes zero
(note that $  \langle  {{ \dot \rho_{\rm{s}}^z}  } \rangle :=  \langle  {{\partial _t  \rho_{\rm{s}}^z}  } \rangle  = \partial _t  \langle  {{  \rho_{\rm{s}}^z}  } \rangle $);
 \begin{eqnarray}
\overline{  \langle  {{ \dot \rho_{\rm{s}}^z}  } \rangle }=0.
 \label{eqn:qsp14}
 \end{eqnarray}
As the result,
the spin continuity equation for conduction electrons, eq. (\ref{eqn:qsp11}),  reads
$ \overline{\langle \nabla \cdot {\mathbf{j}}_{\rm{s}}^{z} \rangle }  =  \overline{ \langle {\mathcal{T}}_{\rm{s}}^z  \rangle }$.
Here it should be noted that 
conduction electrons cannot enter the ferromagnet,\cite{spinwave}   which is an insulator (see Fig. \ref{fig:pumping} (b)).
Consequently by integrating over the volume of the interface and 
adopting the Gauss's divergence theorem,
the time-average of the net spin current pumped into the adjacent non-magnetic metal 
(i.e. $ \int {\mathbf{j}}_{\rm{s}}^{z} \cdot  d{\mathbf{S}}_{\rm{interface}} $ with the surface of the interface $  {\mathbf{S}}_{\rm{interface}}$)  
can be evaluated as 
 \begin{eqnarray}
\overline{\Big{\langle} \int {\mathbf{j}}_{\rm{s}}^{z} \cdot  d{\mathbf{S}}_{\rm{interface}}   \Big{\rangle} }  
&=&   \int_{{\mathbf{x}}\in \rm{(interface)}} \ d{\mathbf{x}} \overline{ \langle  {\mathcal{T}}_{\rm{s}}^z   \rangle }.   
 \label{eqn:qsp15}
 \end{eqnarray}
Experimentally,
this pumped spin current can be detected via the inverse spin Hall effect\cite{ISHE1} in the non-magnetic metal.

Let us emphasize that
the time-average of the pumped net spin current,
$ \overline{{\langle} \int {\mathbf{j}}_{\rm{s}}^{z} \cdot  d{\mathbf{S}}_{\rm{interface}}   {\rangle} }   $,
is expressed only in terms of  the SRT (see eq. (\ref{eqn:qsp15}));
note that  calculating $ \overline{  \langle  {{ \dot \rho_{\rm{s}}^z}  } \rangle }$
has no relation with evaluating the pumped net spin current
even when the total spin angular momentum is conserved.
That is,
the spin density for conduction electrons, $ \rho_{\rm{s}}^z$, is not relevant to quantum spin pumping mediated by magnon.\cite{kurebayashi}
This is one of the main results from our formalism.
Thus from now on, we focus on  $ {\mathcal{T}}_{\rm{s}}^z$  and 
qualitatively  clarify the features of  quantum spin pumping  mediated by magnons. 

\subsection{Spin relaxation torque}
\label{subsec:srt}

It is also easy to see that 
the expectation value of the SRT reads\cite{QSP}
$\langle {\mathcal{T}}_{\rm{s}}^z \rangle  =  \sum_{n=0, \pm 1} \langle  {\mathcal{T}}_{\rm{s}}^{z} (n)  \rangle  \   {\rm{e}}^{2in\Omega t}  + {\cal{O}}(\Gamma ^4)$, 
where $ {\mathcal{T}}_{\rm{s}}^{z} (n) $ represents the (time-independent) expansion coefficient of each angular frequency mode.
Thus the time-average becomes
\begin{eqnarray}
\overline{\langle {\mathcal{T}}_{\rm{s}}^z \rangle } =   \langle  {\mathcal{T}}_{\rm{s}}^{z} (n=0)  \rangle.  
 \label{eqn:qsp17}
\end{eqnarray}

The interface  is,  in general,   a weak coupling regime;\cite{AndoPumping}
the exchange interaction $J$ (see Hamiltonian (\ref{eqn:qsp4})) is supposed to be smaller than the Fermi energy 
and the exchange interaction among ferromagnets.
In addition, a weak transverse magnetic field $\Gamma $ is applied.
Therefore we are allowed to treat $ { \cal{H}}_{\rm{ex}}^{\prime} $,   $V_{\rm{el}}^{\Gamma } $, and   $V_{\rm{mag}}^{\Gamma } $
as perturbative terms to evaluate the SRT.

Through the standard procedure of the Schwinger-Keldysh (or contour-ordered) Green's function\cite{ramer,kamenev,kita}
and the Langreth method,\cite{haug,tatara,new,rammer}
the SRT,  $  \langle  {\mathcal{T}}_{\rm{s}}^{z} (n=0)  \rangle$, is evaluated as follows
(see also Fig. \ref{fig:EffectiveMag} (b). The detail of the calculation had  been shown in our previous work.\cite{QSP});
\begin{eqnarray}
 \langle {\mathcal{T}}_{\rm{s}}^z (n=0) \rangle  &=  &                     \frac{J }{2}         (\frac{\Gamma _0}{2})^2   S  
                                                                \Big[1-\frac{i}{\tilde S }     \int \frac{ d{ {\mathbf{k}}^{\prime}}}{ (2\pi)^3}   \int \frac{d{\omega}^{\prime} }{2\pi} 
                                                                                                                           {\rm{G}}^{\rm{<}}_{{\mathbf{k^{\prime}}}, \omega^{\prime} } \Big]
                                                                                                                   \int \frac{ d  {\mathbf{k}} }{ (2\pi)^3}     \int \frac{d\omega}{2\pi}   \nonumber  \\
                                                    &\times  &            \Big[ ({\rm{G}}^{\rm{a}}_{0, -\Omega }   +  {\rm{G}}^{\rm{r}}_{0, -\Omega }  )   
                                                              (   {\mathcal{G}}^{\rm{t}}_{\downarrow , {\mathbf{k}}, \omega -\Omega } 
                                                                                   {\mathcal{G}}^{\rm{t}}_{\uparrow  , {\mathbf{k}}, \omega  }  
                                                                                      -   {\mathcal{G}}^{\rm{<}}_{\downarrow , {\mathbf{k}}, \omega -\Omega } 
                                                                                     {\mathcal{G}}^{\rm{>}}_{\uparrow  , {\mathbf{k}}, \omega  })       \nonumber    \\
                                                  &+   &                      ( {\rm{G}}^{\rm{a}}_{0, \Omega }  +  {\rm{G}}^{\rm{r}}_{0, \Omega }  )  
                                                                 (  {\mathcal{G}}^{\rm{t}}_{\downarrow , {\mathbf{k}}, \omega +\Omega } 
                                                                                        {\mathcal{G}}^{\rm{t}}_{\uparrow  , {\mathbf{k}}, \omega  }  
                                                                                         -    {\mathcal{G}}^{\rm{<}}_{\downarrow , {\mathbf{k}}, \omega +\Omega } 
                                                                                           {\mathcal{G}}^{\rm{>}}_{\uparrow  , {\mathbf{k}}, \omega  }  )  \nonumber \\
                                                  &-  &                        ({\rm{G}}^{\rm{r}}_{0, -\Omega }  +{\rm{G}}^{\rm{a}}_{0, -\Omega } )   
                                                                       (  {\mathcal{G}}^{\rm{t}}_{\uparrow , {\mathbf{k}}, \omega -\Omega } 
                                                                                           {\mathcal{G}}^{\rm{t}}_{\downarrow  , {\mathbf{k}}, \omega  }  
                                                                                           -   {\mathcal{G}}^{\rm{<}}_{\uparrow , {\mathbf{k}}, \omega -\Omega } 
                                                                                          {\mathcal{G}}^{\rm{>}}_{\downarrow  , {\mathbf{k}}, \omega  } )   \nonumber  \\
                                                 &- &                           ( {\rm{G}}^{\rm{r}}_{0, \Omega }   +{\rm{G}}^{\rm{a}}_{0, \Omega } )  
                                                                  (      {\mathcal{G}}^{\rm{t}}_{\uparrow , {\mathbf{k}}, \omega +\Omega } 
                                                                                            {\mathcal{G}}^{\rm{t}}_{\downarrow  , {\mathbf{k}}, \omega  }  
                                                                                          -    {\mathcal{G}}^{\rm{<}}_{\uparrow , {\mathbf{k}}, \omega +\Omega } 
                                                                                           {\mathcal{G}}^{\rm{>}}_{\downarrow  , {\mathbf{k}}, \omega  } )\Big]    \nonumber   \\
                                                 &+&    {\cal{O}}(J^0)  + {\cal{O}}(J^2)   + {\cal{O}}(\Gamma ^4)+   {\cal{O}}(JS^{-1}).
 \label{eqn:qsp18}
\end{eqnarray}
The variables $ {\mathcal{G}}^{{\rm{t}}  ({\rm{r}}, {\rm{a}}, {\rm{<}}, {\rm{>}} )  }  $/$ {\rm{G}}^{{\rm{t}}  ({\rm{r}}, {\rm{a}}, {\rm{<}}, {\rm{>}})   }  $
are the fermionic/bosonic time-ordered, retarded, advanced, lesser, and greater Green's functions, respectively.\cite{QSP}
We here have  taken the extended time (i.e. the contour variable) 
defined on the  Schwinger-Keldysh closed time path,\cite{rammer,kita,tatara,haug,new} c, 
on the forward path ${\rm{c}}_{\rightarrow }$; 
$ c =  {\rm{c}}_{\rightarrow } + {\rm{c}}_{\leftarrow } $.
Even when the time is located  on the backward path ${\rm{c}}_{\leftarrow }$, 
the result of the calculation does not change
because each Green's function is not independent;\cite{kita,kamenev,tatara,new}  
$ {\mathcal{G}}^{\rm{r}}  - {\mathcal{G}}^{\rm{a}}  = {\mathcal{G}}^{\rm{>}}  - {\mathcal{G}}^{\rm{<}} $.
This relation comes into effect also in the bosonic case.\cite{kita,kamenev,tatara,new}  

The SRT (eq. (\ref{eqn:qsp18})) has become proportional to $\Gamma _0^2$; 
$ \langle {\mathcal{T}}_{\rm{s}}^z (n=0) \rangle  \propto \Gamma _0^2   $.
Thus the SRT (i.e. the pumped net spin current) can be interpreted as the non-linear response to
the applied time-dependent transverse magnetic field (i.e. quantum fluctuations).
This is  one of the main features of our quantum spin pumping theory;
our theory well describes the experimental features of quantum spin pumping 
that quantum fluctuations are essential.\cite{QSP}

Now, let us introduce the dimensionless SRT, $ {\overline{    \langle   {\mathcal{T}}_{\rm{s}}^z    \rangle}}/ \Lambda $, 
and the one in the wavenumber-space  for conduction electrons, $\langle \widetilde{{\mathcal{T}}_{{\rm{s}}  }^{z}}(n=0) \rangle  $, as follows;
\begin{eqnarray}
     \overline{    \langle   {\mathcal{T}}_{\rm{s}}^z    \rangle}   &:= &       \Lambda        \int_0^{\infty }    ( \sqrt{\frac{F}{\epsilon _{\rm{F}}}} dk) \    
                                                                        \langle    \widetilde{{\mathcal{T}}_{{\rm{s}}  }^{z}}(n=0)   \rangle        \label{eqn:qsp19-2}                    \\   
            & {\rm{with}}  &   \nonumber   \\   
      \Lambda   &:=&        \frac{\sqrt{\epsilon _{\rm{F}}} S  {{ {\Gamma} _0}}^2}{4(2\pi \sqrt{F})^3  }.  
 \label{eqn:qsp19}
\end{eqnarray}
We have denoted  as $F:= (2m_{\rm{el}})^{-1} $
and the parameters, $\epsilon _{\rm{F}}$ and $k$, 
represent the Fermi energy and the wavenumber of conduction electrons.

According to Hamiltonian (\ref{eqn:qsp5}) and (\ref{eqn:qsp01}),
the effective magnetic field along the quantization axis for conduction electrons, $s^z= c^{\dagger }\sigma ^z c/2$,  reads $2JS+B $
and that for magnons does $B$ at the interface.
On top of this,
the applied time-dependent transverse magnetic field $\Gamma(t)=\Gamma _0 {\rm{cos}}(\Omega t) $ (i.e. quantum fluctuations) 
affects conduction electrons as well as magnons at the interface; Fig. \ref{fig:pumping}.
Thus,
the SRT (eq. (\ref{eqn:qsp18})) becomes a non-zero value around the points\cite{QSP} $ \Omega = 2JS+B $ and $\Omega =B$,
which are generated by ESR and FMR, respectively.
That is (eq. (\ref{eqn:qsp15})),
spin pumping is generated by ESR ($ \Omega = 2JS+B $) as well as FMR ($ \Omega = B $).\footnote{
Our  theory based on Schwinger-Keldysh formalism
describes spin pumping by ESR and FMR;
see Fig. \ref{fig:FMRESR}.
}

As you know,
concerning spin pumping by FMR ($\Omega =B$),
Tserkovnyak et al. \footnote{
Concerning the distinction between the pioneering theory proposed by Tserkovnyak et al.
and our quantum spin pumping theory based on Schwinger-Keldysh formalism,
please see our preceding paper [K. Nakata. J. Phys. Soc. Jpn., ${\mathbf{81}}$ (2012) 064717 / arXiv:1201.1947], 
which has already discussed the issue in detail.\cite{QSP}
} 
have already  revealed the dynamics
and have given clear explanations.\cite{mod2,battery,bauerreview} 
Therefore from now on,
we exclusively focus on quantum spin pumping by  ESR ($ \Omega = 2JS+B $).

\begin{figure}[b]
\begin{center}
\includegraphics[width=8cm,clip]{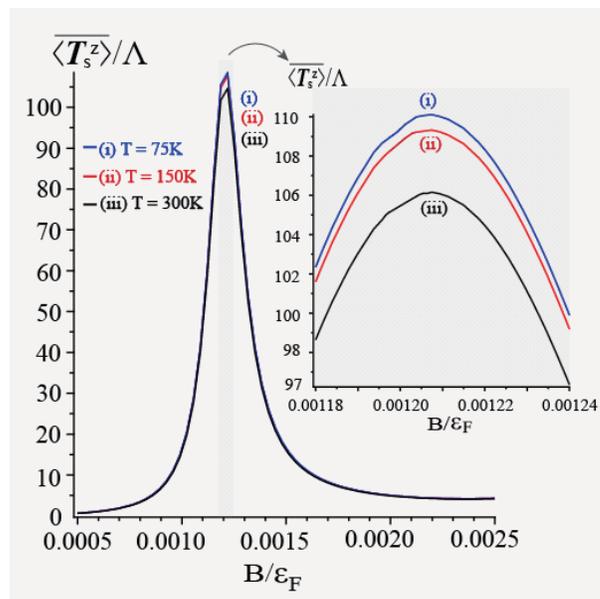}
\caption{
The temperature dependence of the SRT around the ESR point ($ \Omega = 2JS +B$).
When temperature rises,  the SRT  becomes smaller.
As a typical case,\cite{QSP} each parameter is set as follows;\cite{AndoPumping,xiao,spinwave,kittel,QSP}
$ \epsilon _{\rm{F}} = 5.6  $ eV,
$  T=300 $ K,
$ F:= (2m_{\rm{el}})^{-1}=4$ eV {\AA}$^2$,
$ D:= (2m)^{-1}=0.3$ eV {\AA}$^2$,
$a_0 =3$  {\AA},
$S=1/2$,
$ \Omega /{\epsilon _{\rm{F}}} =  0.0032$, and
$  J /{\epsilon _{\rm{F}}} =  0.002 $.
 \label{fig:ESRTemp} }
\end{center}
\end{figure}

\section{Temperature dependence of  quantum spin pumping by ESR
}
\label{sec:temp}

From now on,
we  investigate the features of  quantum spin pumping mediated by magnons under  ESR ($\Omega =2JS+B$),
in particular the temperature dependence.

Fig. \ref{fig:ESRTemp}
shows the temperature dependence of the SRT  by ESR;
it is clear  that when temperature rises, 
the SRT  becomes smaller.
To find the reason and the microscopic origin is the main aim of this section (sec. \ref{subsec:effective} and   \ref{subsec:split}).
The resulting features of quantum spin pumping by ESR is shown in sec. \ref{subsec:ele}.

Here, it would be helpful to remark that our present theory is applicable to the quantum spin pumping systems where the following conditions, (i) and (ii),  are satisfied;
(i) the magnitude of the applied transverse magnetic field $\mid \Gamma   \mid $ and the exchange interaction between conduction electrons and magnons
    are smaller than the Fermi energy and the exchange interaction among ferromagnets,
(ii) temperature of the system is lower than Fermi temperature and Curie temperature.

\subsection{Effective magnetic field for conduction electrons}
\label{subsec:effective}

As mentioned (Hamiltonian (\ref{eqn:qsp4-3})),
localized spins $S$ act as  effective magnetic fields along the quantization axis for conduction electrons.
According to eq. (\ref{eqn:qsp18}), 
the effective magnetic field  at finite temperature  $  S_{\rm{eff}}(T)  $ becomes
\begin{eqnarray}
S_{\rm{eff}}(T)   &=&   S  \Big[1-\frac{i}{\tilde S }   \int \frac{ d{ {\mathbf{k}}^{\prime}}}{ (2\pi)^3}    \int \frac{d{\omega}^{\prime} }{2\pi} 
                                             {\rm{G}}^{\rm{<}}_{{\mathbf{k^{\prime}}}, \omega^{\prime} } \Big],    
 \label{eqn:qsp29}                                                                   
\end{eqnarray}
and it is the monotone decreasing function in respect to temperature $T$ 
around the ESR point
($\Omega =2JS+B $, Fig. \ref{fig:EffectiveMag}  (b-iii));
 $   d S_{\rm{eff}}  (T) /(dT) < 0$.             
In addition,
the SRT is proportional to   $  S_{\rm{eff}}(T)  $ (see eq. (\ref{eqn:qsp18}));
 $ \langle {\mathcal{T}}_{\rm{s}}^z (n=0) \rangle  \propto    J  S_{\rm{eff}}(T)      \Gamma _0^2$.     
As the result,  the SRT by ESR (see Fig. {\ref{fig:ESRTemp}}. See also sec. \ref{subsec:ele} in advance)
is also the monotone decreasing function in respect to temperature $T$;
  $  d \langle {\mathcal{T}}_{\rm{s}}^z (n=0) \rangle /(dT)  < 0$.           
This behavior of quantum spin pumping by ESR ($\Omega =2JS+B$) at finite temperature
is the main distinction from that of the standard spin pumping  by FMR ($\Omega =B$).\cite{AndoPumping,andoprivate,czeschka}

\subsection{Three-magnon splittings}
\label{subsec:split}

\begin{figure}[b]
\begin{center}
\includegraphics[width=12cm,clip]{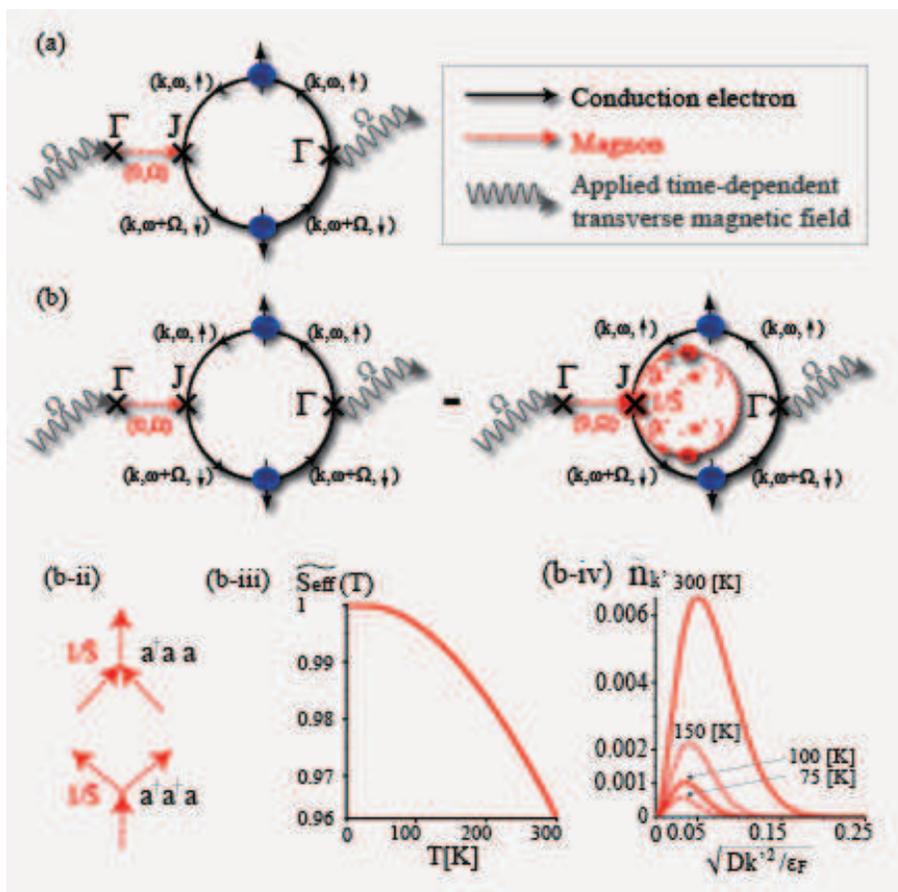}
\caption{
(a)
A Feynman diagram of the SRT in the  large-$S$ limit.
(b) 
A Feynman diagram of the SRT with three-magnon splittings;
$  \langle {\mathcal{T}}_{\rm{s}}^z (n=0) \rangle$.
(b-ii) 
The schematic picture of  three-magnon splittings.
(b-iii) 
The plot of  the function
$   \widetilde{{S}_{\rm{eff}}}(T) :=   S_{\rm{eff}}(T)/S     $
as a function of temperature $T$
on the ESR point ($\Omega =2JS+B$).
(b-iv)
The plot of the dimensionless distribution function of magnons 
in the dimensionless wavenumber space  
on the ESR point ($\Omega =2JS+B$);
${\widetilde{n}}_{k^{\prime}}$.
 \label{fig:EffectiveMag} }
\end{center}
\end{figure}

We microscopically go after the origin of the effective magnetic field at finite temperature  $  S_{\rm{eff}}(T)  $.
As shown in eqs. (\ref{eqn:qsp3-2}) and (\ref{eqn:qsp3-2-2}),
we have rewritten the localized spin degrees of freedom into magnon ones via the Holstein-Primakoff transformation;
$S^+({\mathbf{x}},t)
= \sqrt{2\tilde S} [1- {a^\dagger({\mathbf{x}},t) a({\mathbf{x}},t)}/({4\tilde S})]a({\mathbf{x}},t)    
+{\cal{O}}({\tilde S}^{-3/2})
=  (S^{-})^{\dagger }$.   
Note that we have expanded up to $ {\cal{O}}({\tilde S}^{-1/2}) $
and have included the effects of  three-magnon splittings\cite{kurebayashi,andoprivate} (Fig. \ref{fig:EffectiveMag}  (b-ii));
\begin{eqnarray}
\frac{a^\dagger({\mathbf{x}},t)   a({\mathbf{x}},t)  a({\mathbf{x}},t)}{\tilde S}     \     \  {\rm{and}}    \     \
 \frac{a^\dagger({\mathbf{x}},t)   a^\dagger({\mathbf{x}},t)   a({\mathbf{x}},t)}{\tilde S},
 \label{eqn:qsp32}
\end{eqnarray}            
which lead to the loop effects\cite{peskin} (i.e. quantum effects) expressed  as
(see eq. (\ref{eqn:qsp29}) and  Fig. \ref{fig:EffectiveMag} (b)) 
\begin{eqnarray}
   S  \Big[-\frac{i}{\tilde S }   \int \frac{ d{ {\mathbf{k}}^{\prime}}}{ (2\pi)^3}    \int \frac{d{\omega}^{\prime} }{2\pi} 
                                             {\rm{G}}^{\rm{<}}_{{\mathbf{k^{\prime}}}, \omega^{\prime} } \Big]  
 =: - \Lambda ^{\prime}   \int_0^{\infty }    ( \sqrt{\frac{D}{\epsilon _{\rm{F}}}} dk^{\prime})    \   
                                                                                  {\widetilde{n}}_{k^{\prime}}   
       \label{eqn:qsp49}                                                            
\end{eqnarray}
with
\begin{eqnarray}
 \Lambda ^{\prime}    &:=& \frac{S}{4 {\pi}^3  \tilde S}  \Big( \frac{\epsilon _{\rm{F}}}{D} \Big)^{3/2}.
 \label{eqn:qsp50}   
\end{eqnarray}
We have denoted  as $D:= (2m)^{-1} $
and the variable ${\widetilde{n}}_{k^{\prime}}$ in eq. (\ref{eqn:qsp49}) represents
the dimensionless distribution function of magnons in the dimensionless wavenumber space for magnons 
($   \sqrt{{D}/{\epsilon _{\rm{F}}}} dk^{\prime} $); Fig. \ref{fig:EffectiveMag} (b-iv).
It is apparent from Fig. \ref{fig:EffectiveMag} (b-iv) and eqs.  (\ref{eqn:qsp29}) and (\ref{eqn:qsp49})   that
the three-magnon splittings excite the non-zero modes of magnons (${\mathbf{k}}^{\prime} \not=0 $);
when temperature rises,
the  wavenumber of excited magnons becomes larger
and the number of  excited magnons also  increases.
Consequently,
the magnitude of localized spins along the quantization axis become smaller
and it leads to the behavior of quantum spin pumping by ESR at finite temperature which is discussed in sec. \ref{subsec:effective}.
That is,
the three-magnon splittings characterize the effective magnetic field at finite temperature $  S_{\rm{eff}}(T)  $
and the temperature dependence of quantum spin pumping by ESR.

If we expand up to only $ {\cal{O}}(\sqrt{{\tilde S}})$;
$S^+({\mathbf{x}},t)
= \sqrt{2\tilde S} a({\mathbf{x}},t)    +{\cal{O}}({\tilde S}^{-1/2})  
= (S^{-})^{\dagger }$,
and neglect three-magnon splittings,
the effective magnetic field is reduced to $ S$ (Fig. \ref{fig:EffectiveMag} (a));
$S_{\rm{eff}}   \rightarrow   S$.   
This corresponds to the large-$S$ limit.\cite{lieb}
In this case,
only the zero mode of magnons is relevant to the SRT (see eqs. (\ref{eqn:qsp18}) and (\ref{eqn:qsp49}))
and non-zero modes of magnons are not excited.

\subsection{Contribution of conduction electrons}
\label{subsec:ele}

The SRT at finite temperature, $  \overline{\langle {\mathcal{T}}_{\rm{s}}^z  \rangle }\mid _{T}  =  \langle {\mathcal{T}}_{\rm{s}}^z (n=0) \rangle$, 
can be expressed also as follows (see eqs. (\ref{eqn:qsp18}) and (\ref{eqn:qsp29})); 
\begin{eqnarray}
\overline{\langle {\mathcal{T}}_{\rm{s}}^z  \rangle }\mid _{T} &=&
{\overline{\langle {\mathcal{T}}_{\rm{s}}^z  \rangle}\mid _{T=300 [\rm{K}]}} \times      \eta ^{\rm{ratio}} (T)  \times     {{S}_{\rm{eff}}^{\rm{ratio}}}(T)     \\
 \label{eqn:qsp37}       
                 & {\rm{with}} &   \nonumber   \\
\eta ^{\rm{ratio}} (T)  &:=&  
\frac{ {{\mathcal{T}}_{\rm{s}}^{z{\mathchar`-}\rm{ratio}}} (T)}{{{S}_{\rm{eff}}^{\rm{ratio}}}(T)},
\label{eqn:qsp36}          \\
 {{\mathcal{T}}_{\rm{s}}^{z{\mathchar`-}\rm{ratio}}} (T)  &:=&
\frac{\overline{\langle {\mathcal{T}}_{\rm{s}}^z  \rangle }\mid _{T}}{\overline{\langle {\mathcal{T}}_{\rm{s}}^z  \rangle}\mid _{T=300 [\rm{K}]}},   
\label{eqn:qsp36-2}    \\
 &   {\rm{and}}&   \nonumber \\
{{S}_{\rm{eff}}^{\rm{ratio}}}(T)  &:=&  
  \frac{ \widetilde{{S}_{\rm{eff}}}(T) }{ \widetilde{{S}_{\rm{eff}}}(T=300 [\rm{K}]) }.                    
 \label{eqn:qsp36-3}       
\end{eqnarray}         
By using  eq. (\ref{eqn:qsp36}),
the dimensionless SRT  at finite temperature can be rewritten as 
${{\mathcal{T}}_{\rm{s}}^{z{\mathchar`-}\rm{ratio}}} (T) =     \eta ^{\rm{ratio}} (T)   \times         {{S}_{\rm{eff}}^{\rm{ratio}}}(T)   $.
Consequently, it is clear that
the variable $  \eta ^{\rm{ratio}} (T)$ represents the contribution of conduction electrons to spin pumping 
(see also eqs. (\ref{eqn:qsp18}) and (\ref{eqn:qsp29}))
and thus, it can be interpreted to
correspond to the mixing conductance in the spin pumping theory proposed by Tserkovnyak et al.\cite{mod2,battery,bauerreview}

\begin{figure}[b]
\begin{center}
\includegraphics[width=9cm,clip]{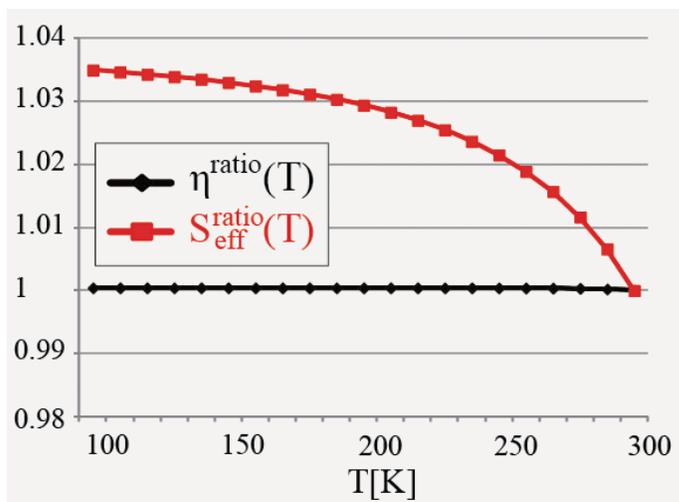}
\caption{
Plot of  $ \eta ^{\rm{ratio}} (T)  $  and $ {{S}_{\rm{eff}}^{\rm{ratio}}}(T) $  as a function of temperature $T$
on the ESR point ($\Omega =2JS+B$).
Compared with  $ {{S}_{\rm{eff}}^{\rm{ratio}}}(T) $,
the function $  \eta ^{\rm{ratio}} (T)$ is little influenced by temperature; $  \eta ^{\rm{ratio}} (T) \sim 1$.
 \label{fig:ESRT} }
\end{center}
\end{figure}

One can easily see from Fig. \ref{fig:ESRT} that $  \eta ^{\rm{ratio}} (T)$ is little influenced by temperature;
\begin{eqnarray}
\eta ^{\rm{ratio}} (T)  \sim 1.
 \label{eqn:qsp38}       
\end{eqnarray}            
This temperature dependence of $ \eta ^{\rm{ratio}} (T)$
qualitatively shows the good agreement with the experimental result by Czeschka et al.\cite{czeschka}
 (i.e. the measurement of the mixing conductance under the standard spin pumping by  FMR)
 that the mixing conductance is little influenced by temperature.
 That is,
 this temperature dependence (eq. (\ref{eqn:qsp38})) is the common properties of spin pumping by FMR and ESR.

In conclusion,
the temperature dependence of quantum  spin pumping by ESR
is determined mainly  by  $ S_{\rm{eff}}(T) $,
which is governed by three-magnon splittings.
On top of this,
 $  \eta ^{\rm{ratio}} (T)$,  which represents the contribution of conduction electrons to spin pumping 
 and corresponds to the mixing conductance in the spin pumping theory proposed by Tserkovnyak et al.,
is little influenced by temperature.
This qualitatively shows the good correspondence with the experiment by Czeschka et al.;\cite{czeschka} 
this temperature dependence (eq. (\ref{eqn:qsp38})) is the common properties of spin pumping by FMR and ESR.

\section{Summary and discussion}
\label{sec:summary}

We have clarified the temperature dependence of quantum spin pumping generated by ESR
and have found the microscopic origin.
When temperature rises, the pumped net spin current  under ESR decreases;
this is our theoretical prediction.
This temperature dependence is governed by three-magnon splittings,
which excite non-zero modes of magnons.
On top of this,
 $  \eta ^{\rm{ratio}} (T)$,  which represents the contribution of conduction electrons to spin pumping 
 and corresponds to the mixing conductance in the spin pumping theory proposed by Tserkovnyak et al.,
is little influenced by temperature.
This qualitatively shows the good correspondence with the experiment by Czeschka et al.\cite{czeschka} 
That is,
the temperature dependence (i.e. $  \eta ^{\rm{ratio}} (T) \sim 1$) is the common properties of spin pumping by FMR and ESR.

Let us remark that we have theoretically predicted that 
the pumped net spin current  by ESR decreases when temperature rises;
this temperature dependence of quantum spin pumping by ESR will be  
experimentally  confirmed  by the inverse spin Hall effect.\cite{ISHE1}
Although external pumping magnetic fields are supposed to be applied to the whole of the sample as well as the interface 
due to the restriction of experimental techniques\cite{Bauer}  (see Fig. \ref{fig:pumping}),
fortunately  only the ESR at the interface occurs
when the angular frequency is tuned to $\Omega =2JS+B $.
Other resonances take place in other regime;
ESR at the non-magnetic metal  and FMR at the interface and the ferromagnetic metal occur when $\Omega =B $.
Thus by adjusting the angular frequency of the applied magnetic field,
the temperature dependence of the pumped spin current purely by ESR at the interface will be observable. 

On the other hand,
to clarify the effects of the unusual energy dispersion of the lowest magnon mode in YIG,
which is a relevant material to the experiment of magnon BEC\cite{demokritov,demidov,chumak} and spin pumping,\cite{spinwave,sandweg,kurebayashi}
is a significant theoretical issue.

\ack
We  would like to thank 
G. E. Bauer for invaluable discussion and significant comments on our quantum spin pumping theory
from the viewpoint of experiments
during the 4th International Workshop on Spincaloritronics held at Institute for Materials Research of Tohoku University;
he gave us the potential opportunity to attend the conference. 
We  also thank  K. Ando for sending the invaluable presentation file, prior to the publication, of the 6th International School and Conference on Spintronics
and Quantum Information Technology ({\textit{Spintech6}}).  
Last,
we would like to thank   the members of the Condensed Matter Theory and Quantum Computing Group of the University of Basel
for the warm hospitality and financial support 
during our stay under the young researchers exchange program by the Yukawa Institute for Theoretical Physics.
In particular, we are grateful to Dr. Kevin A. van Hoogdalem and Prof. Daniel Loss for significant discussion during our stay.

We are supported by the Grant-in-Aid for the Global COE Program
"The Next Generation of Physics, Spun from Universality and Emergence"
from the Ministry of Education, Culture, Sports, Science and Technology (MEXT) of Japan.

\appendix

\section{Quantum spin pumping by FMR and ESR}
\label{sec:FMRESR}

\begin{figure}[h]
\begin{center}
\includegraphics[width=10cm,clip]{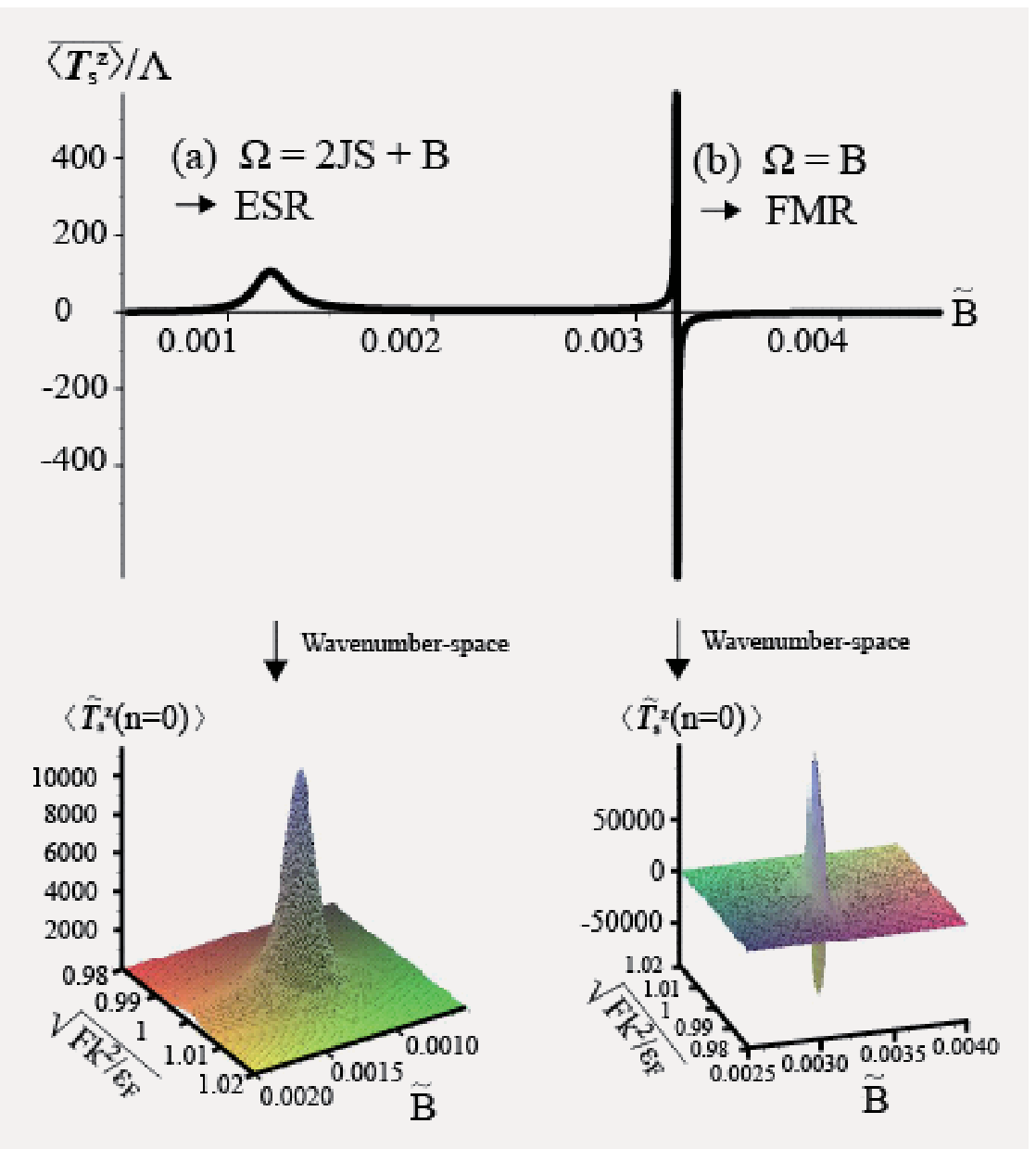}
\caption{
Plot of the SRT as a function of ${\tilde B}:=B/{\epsilon _{\rm{F}}}$; 
$  \overline{    \langle   {\mathcal{T}}_{\rm{s}}^z    \rangle} /\Lambda  $ and $    \langle    \widetilde{{\mathcal{T}}_{{\rm{s}}  }^{z}}(n=0)   \rangle $.
Sharp peaks exist on the point; (a) $ \Omega = 2JS+B $ and  (b) $\Omega =B$,
which has resulted from (a) ESR and (b) FMR.
{\textbf{ {\textit{Note};
for reader's convenience,
this picture (or appendix) has been newly added into the manuscript submitted to arXiv.
}  }
}
 \label{fig:FMRESR} }
\end{center}
\end{figure}

Fig. \ref{fig:FMRESR} shows $\overline{    \langle   {\mathcal{T}}_{\rm{s}}^z    \rangle} /\Lambda  $ and  
$  \langle    \widetilde{{\mathcal{T}}_{{\rm{s}}  }^{z}}(n=0)   \rangle    $. 
It is clear that sharp peaks exist on the point\cite{andoprivate} (a) $ \Omega = 2JS+B $ and  (b) $\Omega =B$.
The effective magnetic field along the quantization axis for conduction electrons reads $2JS+B $
and that for magnons (i.e. localized spins) does $B$.
Therefore it can be concluded that 
the sharp peak on the point (a) $ \Omega = 2JS+B $ has originated from ESR
and that on the point (b) $\Omega =B$ from FMR.
This is the natural result from the fact that at the interface
quantum fluctuations (i.e. time-dependent transverse magnetic fields)
affect conduction electrons as well as localized spins (i.e. magnons)
which is acting as  effective magnetic fields for conduction electrons; $2JS$.

Last, let us stress that
although in the present manuscript
we have explicitly clarified that
the mixing conductance under spin pumping by ESR is little influenced by temperature,
our quantum spin pumping theory also shows that
the mixing conductance under spin pumping by FMR\cite{czeschka} is little influenced by temperature;\footnote{
K. Nakata, unpublished.
}
this theoretical result agrees with the experiment by Czeschka et al.
[F. D. Czeschka et al., Phys. Rev. Lett., {$\mathbf{107}$}, 046601 (2011)].

\section*{References}
\bibliographystyle{unsrt}
\bibliography{PumpingRef}

\end{document}